# Predicting Affective States from Screen Text Sentiment

Songyan Teng
songyan.teng@student.unimelb.edu.au
University of Melbourne
Melbourne, Australia

Tianyi Zhang
t.zhang59@student.unimelb.edu.au
University of Melbourne
Melbourne, Australia

Simon D'Alfonso
dalfonso@unimelb.edu.au
University of Melbourne
Melbourne, Australia

Vassilis Kostakos
vassilis.kostakos@unimelb.edu.au
University of Melbourne
Melbourne, Australia

## ABSTRACT

The proliferation of mobile sensing technologies has enabled the study of various physiological and behavioural phenomena through unobtrusive data collection from smartphone sensors. This approach offers real-time insights into individuals' physical and mental states, creating opportunities for personalised treatment and interventions. However, the potential of analysing the textual content viewed on smartphones to predict affective states remains underexplored. To better understand how the screen text that users are exposed to and interact with can influence their affects, we investigated a subset of data obtained from a digital phenotyping study of Australian university students conducted in 2023. We employed linear regression, zero-shot, and multi-shot prompting using a large language model (LLM) to analyse relationships between screen text and affective states. Our findings indicate that multi-shot prompting substantially outperforms both linear regression and zero-shot prompting, highlighting the importance of context in affect prediction. We discuss the value of incorporating textual and sentiment data for improving affect prediction, providing a basis for future advancements in understanding smartphone use and wellbeing.

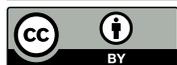

## 1 INTRODUCTION

Mobile sensing technologies have been widely used in wellbeing studies and applications, and the significant advancements in sensing over the last decade have spurred heightened interest in this field, often referred to as "digital phenotyping". This approach often involves the use of smartphone sensors to continuously and unobtrusively collect data on various physiological and behavioural phenomena [9]. Data from a range of smartphone sensors can be integrated to obtain a comprehensive understanding of a person's surroundings, activities, and behaviours [1]. This approach allows for real-time monitoring and analysis of individuals' physical and mental states, providing valuable insights into their overall wellbeing and creating opportunities for delivering recommendations and interventions based on the user's context. For instance, Wang et al. [22] used smartphone sensing data to assess stress and academic performance among college students, highlighting the potential of mobile sensing to capture detailed behavioural data and its correlation with mental health. Similarly, sensor data can also be analysed to detect physical activity [23] and sleep [10]. By analysing data collected from smartphone sensors such as accelerometers, GPS, and phone calls, digital phenotyping can be used to quantify wellbeing metrics and predict human behaviour. For example, frequent location changes and irregular activity patterns captured through accelerometer and GPS data can signal symptoms of bipolar disorder [5], while reduced mobility and social interaction can be indicative of depression [17].

In addition to data captured from the physical environment, textual content from smartphones has been studied to understand wellbeing. Textual data, including messages, social media interactions, and web browsing history, can provide rich insights into an individual's thoughts, emotional state, social interactions, and cognitive patterns [20]. Textual content analysis involves examining the language and sentiment expressed in written communication to identify indicators of mental health. For instance, studies have shown that the use of certain words and phrases in texting can be indicative of depression, anxiety, or other mental health conditions [16]. Similarly, De Choudhury et al. [3] analysed Twitter posts to identify users at risk of depression, finding that language patterns, such as the use of negative affect words, were predictive of depressive symptoms.

One approach for capturing textual data is to collect keystroke data from the smartphone keyboard [4], which can provide insights into wellbeing such as emotion [14] and stress [12]. However, this approach does not capture all the text content smartphone users are exposed to, and therefore lacks the contextual information with which users are engaged. To tackle this, an alternative approach is to capture screenshots from mobile devices with high frequency [11]. These screenshots – or Screenomes – are typically taken every few seconds, providing a detailed visual record of individuals' smartphone usage. This technique enables the analysis of broad aspects of smartphone use, but has multiple limitations including difficulty of annotating screenshots, large storage requirements, and intermittent data collection. Teng et al. [18] introduced the screen text sensor, aimed to capture textual content on smartphones using Android's accessibility services. This information can potentially be used to shed light on relationships between interactions with textual content on smartphones and the user's wellbeing.

To better understand how the textual content that smartphone users interact with can influence their wellbeing, we analysed a subset of data obtained from a 2023 semester-long (17-week) field study of approximately 150 university students [2]. This study aimed to capture both their smartphone usage, specifically their screen text, and their affective states in their real-life context. In this paper, we explore how an individual's screen text can be used as predictors of affect and how this can be achieved using text-analysis methods. We discuss the implications of our findings and how they can be extended in developing methods to better understand smartphone use and wellbeing.

## 2 METHODS

### 2.1 Data Collection

To capture the students' smartphone usage, we used the AWARE-Light sensing application [21] and its screen text sensor [18] to collect textual data that appear on the students' phone screens. At the end of each week, participants were asked to complete the International version of the Positive and Negative Affect Schedule (I-PANAS-SF) questionnaire [19] through an online survey platform. This questionnaire consists of 10 affect ratings (5 positive, 5 negative), each measured on a Likert scale from 1 ("Never") to 5 ("Always").

### 2.2 Experiment Setup

*2.2.1 Data Preprocessing and Feature Generation.* Data collected from the screen text sensor on AWARE-Light is formatted in plain text, where all the text and its on-screen coordinates are stored in a single string and delimited. These coordinates can be used to ensure that each text element is stored in the correct position within the string, relative to the other elements of the string in a top-to-bottom, left-to-right approach. After sorting the texts, we implemented a deduplication algorithm to remove text overlaps between consecutive screens. These overlaps in the screen text data can often occur in interactions such as scrolling due to the sensor capturing all user interactions and changes on the screen [18]. By discarding these overlaps, we ensure that each screen's content is recorded as a single, distinct entry.

We then applied natural language processing (NLP) methods to the text, including word tokenisation, stopword removal, and text stemming to normalise and simplify the text data for further analysis [8]. To compute a sentiment value for each screen of text, we used a sentiment analysis pipeline with DistilBERT [13], which returns probabilistic ratings for positive, neutral, and negative sentiment. From each rating, we compute an aggregate sentiment score by first assigning a weight of +1 to positive sentiments, 0 to neutral sentiments, and −1 to negative sentiments. We then sum the resulting values using the following formula:

$$Score_{sentiment} = Prob_{pos} * 1 + Prob_{neg} * (-1) + Prob_{neu} * 0$$

which condenses multiple sentiment ratings into a single, interpretable value that reflects the balance of positive and negative feelings. A positive or negative aggregate score indicates a predominance of positive or negative sentiment, respectively, while a predominantly neutral sentiment will produce an overall neutral score.

Because students reported their affects weekly, we split their text data into daily subsets, which could then be concatenated to understand weekly wellbeing. For each student, we computed the daily mean sentiment score captured in the textual data they viewed, which can be used to identify relationships between their smartphone use and affective states.

*2.2.2 Analysis Methods.* To investigate the relationship between screen text and affects across different people, we randomly selected two participants who had complete sets of questionnaire data from our study as a small pilot analysis. This allowed for sufficient training to assess the performance and generalisability of our predictive models while maintaining a sizeable evaluation dataset.

In our study, we focused on predicting the levels for each of the 10 affects from the I-PANAS-SF questionnaire based on the sentiment of the screen text. While previous studies using machine learning models to predict psychometric scores have generally focused on the overall score of a questionnaire, the potential for more detailed predictions of individual scale items (in this case aspects of affects and feelings) has seen little research.

To achieve this, we first conducted a series of linear regression analyses using the sentiment scores as predictors. We randomly selected 9 whole weeks of data for training the model and reserved the remaining 8 whole weeks for evaluation. This split was repeated 5 times, each time with a different randomly selected 9-week training dataset, to ensure the robustness and reliability of our results. This methodology ensures that our findings are not contingent upon a single data split.

To further evaluate our approach, we also leveraged a large language model (LLM), Gemini Pro 1.5 [6] in our experiments. We performed zero-shot prompting 5 times, with each run using one of the 8-week evaluation data splits that we used for linear regression. For these prompts, we set the temperature parameter of the model to 0 to ensure deterministic outputs. Zero-shot prompting involves prompting the LLM to make predictions without providing any specific training examples. We tested this approach to understand how well the LLM's general knowledge can predict affective states based on sentiment from screen text.

Additionally, we conducted multi-shot prompting experiments. In this approach, we provided Gemini with 9 weeks of sentiment scores and corresponding affect ratings as training data and then prompted it to predict affect ratings for the remaining 8 weeks. This process was repeated 5 times, using the same training and evaluation splits as the linear regression. We implemented multi-shot prompting as it allows the LLM to leverage specific examples from our dataset, which we believe is largely unseen data for an LLM due to the rich and diverse user-specific text it contains. This form of model learning can potentially improve its prediction accuracy by analysing patterns within the given training data. The prompts that we used are detailed in Section 2.3.

By comparing the results from linear regression, zero-shot prompting, and multi-shot prompting, we aim to evaluate the effectiveness of different predictive approaches between screen text and affect and potentially uncover novel analysis methods for screen text.

## 2.3 Prompt Design

To guide the LLM in producing responses relevant to affect prediction, we designed prompts to utilise sentiment scores from our dataset and make predictions about user affect. We describe the exact prompt we used below for each evaluation method.

**Zero-Shot:** We conducted zero-shot prompting by instructing the LLM to predict the user's affects for each of 8 randomly-selected weeks, given only the sentiments of text viewed each day and no prior knowledge of the dataset. We constructed the prompt as follows:

```
Here are the sentiments of text that a university student has viewed on their smartphone for each day over a week. The sentiments range from -1 to 1, with -1 being most Negative and 1 being most Positive.
Your task is to rate how the student felt for each of the following feelings based on the sentiments of the text they have viewed:

Active
Determined
Attentive
Inspired
Alert
Upset
Hostile
Ashamed
Nervous
Afraid

For each feeling, choose a Likert score ranging from 1 to 5 that best represents how the student generally felt during the week, where 1 represents Never and 5 represents Always.

Sentiments of text the student has viewed on their smartphone over a week:

Day 1: {sentiment of text viewed on the first day of the week}
Day 2: {sentiment of text viewed on the second day of the week}
Day 3: {sentiment of text viewed on the third day of the week}
Day 4: {sentiment of text viewed on the fourth day of the week}
Day 5: {sentiment of text viewed on the fifth day of the week}
Day 6: {sentiment of text viewed on the sixth day of the week}
Day 7: {sentiment of text viewed on the seventh day of the week}

When predicting information for a single week, only consider data from that specific week.
Provide your choices in the following format and nothing else:

Active: [<predicted number>]
Determined: [<predicted number>]
Attentive: [<predicted number>]
Inspired: [<predicted number>]
Alert: [<predicted number>]
Upset: [<predicted number>]
Hostile: [<predicted number>]
Ashamed: [<predicted number>]
Nervous: [<predicted number>]
Afraid: [<predicted number>]
```

**Multi-Shot:** We conducted multi-shot prompting by instructing the LLM to predict the user's affects for each of 8 randomly-selected weeks, given the sentiments of text viewed each day as well as the weekly sentiment scores and user affect ratings for the remaining 9 weeks. We constructed the prompt as follows:

```
You will be given a series of sentiments for texts that a university student has viewed on their smartphone for each day over a week. The sentiments range from -1 to 1, with -1 being most Negative and 1 being most Positive.
You will also be given ratings for how the student felt for each of the following feelings based on the sentiments of the text they have viewed:

Active
Determined
Attentive
Inspired
Alert
Upset
Hostile
Ashamed
Nervous
Afraid

Your task is to identify the relationship between the sentiments of the texts viewed by the student and the student's feelings, and use this relationship to rate the student's feelings based on the sentiments of the texts they have viewed.

### Example 1
  Refer to the "Multi-Shot Example Prompt" section below for the format of each
  example.

{remaining examples}

### Task
Based on what you have learned about the student, rate how the student felt for each of the following feelings based on the sentiments of the text they have viewed:

Active
Determined
Attentive
Inspired
Alert
Upset
Hostile
Ashamed
Nervous
Afraid

For each feeling, choose a Likert score ranging from 1 to 5 that best represents how the student generally felt during the week, where 1 represents Never and 5 represents Always.

Sentiments of text the student has viewed on their smartphone over a week:

Day 1: {sentiment of text viewed on the first day of the week}
Day 2: {sentiment of text viewed on the second day of the week}
Day 3: {sentiment of text viewed on the third day of the week}
Day 4: {sentiment of text viewed on the fourth day of the week}
Day 5: {sentiment of text viewed on the fifth day of the week}
Day 6: {sentiment of text viewed on the sixth day of the week}
Day 7: {sentiment of text viewed on the seventh day of the week}

When predicting information for a single week, only consider data from that specific week.
Provide your choices in the following format and nothing else:

Active: [<predicted number>]
Determined: [<predicted number>]
Attentive: [<predicted number>]
Inspired: [<predicted number>]
Alert: [<predicted number>]
Upset: [<predicted number>]
Hostile: [<predicted number>]
Ashamed: [<predicted number>]
Nervous: [<predicted number>]
Afraid: [<predicted number>]
```

**Multi-Shot Example Format:** For each of the 9 weekly examples in our multi-shot approach, we constructed the example prompt as follows:

```
Here are the sentiments of text that a university student has viewed on their
smartphone for each day over a week.
The sentiments range from -1 to 1, with -1 being most Negative and 1 being most
Positive.

Each feeling is represented using a Likert score ranging from 1 to 5 that
represents how the student generally felt during the week, where 1 represents
Never and 5 represents Always.

Sentiments of text the student has viewed on their smartphone over a week:

Day 1: {sentiment of text viewed on the first day of the week}
Day 2: {sentiment of text viewed on the second day of the week}
Day 3: {sentiment of text viewed on the third day of the week}
Day 4: {sentiment of text viewed on the fourth day of the week}
Day 5: {sentiment of text viewed on the fifth day of the week}
Day 6: {sentiment of text viewed on the sixth day of the week}
Day 7: {sentiment of text viewed on the seventh day of the week}

Feelings of the student over a week:

Active: {how active the student felt over the week}
Determined: {how determined the student felt over the week}
Attentive: {how attentive the student felt over the week}
Inspired: {how inspired the student felt over the week}
Alert: {how alert the student felt over the week}
Upset: {how upset the student felt over the week}
Hostile: {how hostile the student felt over the week}
Ashamed: {how ashamed the student felt over the week}
Nervous: {how nervous the student felt over the week}
Afraid: {how afraid the student felt over the week}
```

## 3  RESULTS AND DISCUSSION

The prediction accuracies for both participants were evaluated using three methods: Linear Regression, Gemini - Zero-Shot, and Gemini - Multi-Shot.

We used the Mean Absolute Error (MAE) as our evaluation metric, which measures the average magnitude of errors in predictions. For each evaluation run, we predicted ratings for each user affect across each of 8 weeks. We conducted 5 runs for each evaluation method and averaged the MAE across each run.

For evaluating the overall prediction accuracy of each affect for one participant, the MAE calculation can be represented as:

$$\text{MAE}_{\text{affect}} = \frac{1}{R} \sum_{i=1}^{R} \left( \frac{1}{N} \sum_{j=1}^{N} |y_{ij} - \hat{y}_{ij}| \right)$$

where $y_{ij}$ and $\hat{y}_{ij}$ represent the actual and predicted affect rating, respectively, for the $j$-th week in the $i$-th run, $N$ is the number of weeks in each evaluation set (which is 8 in this case), and $R$ is the total number of runs (which is 5 in this case).

*3.0.1  Linear Regression.* The linear regression method consistently exhibited the highest MAE values across all affects for both participants, indicating the least accurate predictions. For Participant 1, the MAE for all affects were substantially higher compared to the LLM evaluations, most notably for affects such as "Inspired" (4.78 ± 2.20), "Upset" (3.54 ± 2.47), and "Alert" (3.53 ± 2.41), as seen in Table 1. Noticeably, some affects such as "Nervous" (3.43 ± 3.30) produced large standard deviations, signalling uncertainty in these predictions. These trends were reflected similarly for Participant 2, where linear regression resulted in high MAE values for "Afraid" (3.44 ± 4.14), "Hostile" (2.98 ± 1.51), and "Ashamed" (2.68 ± 0.80), as displayed in Table 2. Except for "Afraid", predictions for Participant

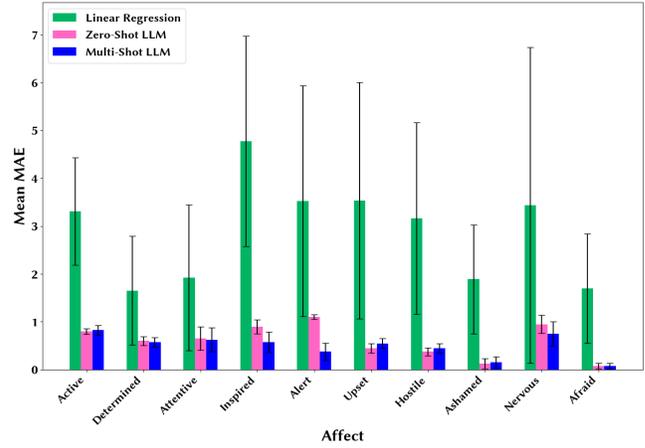

**Figure 1: Participant 1 - Mean MAE for predicting each affect using each evaluation method. Error bars denote standard error.**

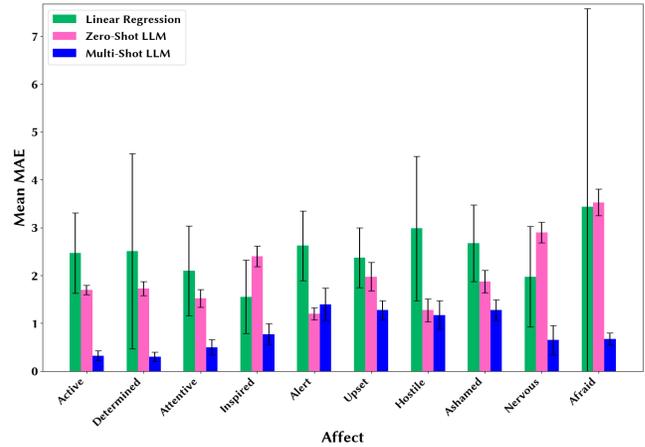

**Figure 2: Participant 2 - Mean MAE for predicting each affect using each evaluation method. Error bars denote standard error.**

2 (Fig. 2) were generally less variant than those for Participant 1 (Fig. 1).

The high MAE values and large standard deviations associated with linear regression can be attributed to several factors. Firstly, linear regression models are inherently simplistic and assume a linear relationship between the input features and the target variable. However, the relationship between screen text sentiment and affective states is likely to be much more complex, which linear regression fails to capture effectively. Secondly, linear regression does not inherently consider the context or nuances in the textual data. Affective states are influenced by subtle cues and context within the text as well as the overall context of when and how this text is captured, which are not adequately captured by a simple linear model. In contrast, the LLM-based methods can be more capable of understanding and leveraging contextual information, leading to more accurate and stable predictions.

*3.0.2 Zero-Shot Prompting.* Zero-shot prompting using Gemini performed significantly better than linear regression, providing lower MAE values for most affects. For Participant 1, the zero-shot approach performed best across all three approaches for "Active" (0.80 ± 0.06), "Upset" (0.45 ± 0.10), "Hostile" (0.38 ± 0.08), and "Ashamed" (0.13 ± 0.11), while achieving the same accuracy as the multi-shot approach for "Afraid" (0.08 ± 0.06), as shown in Table 1. These results indicate a substantial improvement over linear regression. For Participant 2, zero-shot predictions also showed better accuracy than linear regression, though largely performing worse than the multi-shot approach, with only "Alert" (1.20 ± 0.13) being of the highest accuracy, as seen in Table 2. While zero-shot predictions were generally more accurate than linear regression, they were still outperformed by the multi-shot approach in most cases.

The improved performance of the zero-shot approach over linear regression is largely due to the greatly improved capabilities of the model and the additional context it can recognise [15]. Zero-shot prompting leverages the extensive pre-training of the LLM on diverse datasets. This pre-training allows the model to capture complex patterns and relationships in the data, which a simple linear regression model cannot. Consequently, the zero-shot model can make more accurate predictions even without additional task-specific training data. Zero-shot prompting also benefits from the inherent ability of LLMs to understand and process natural language context, which is critical in understanding the text that users engage with. This contextual understanding is crucial for accurately predicting affects, which can help to infer these nuances directly from the input text and lead to more precise predictions compared to linear regression.

However, the zero-shot approach still fell short of the multi-shot approach in most cases. The multi-shot method, which involves providing the model with specific examples and context, further enhances the model's understanding and prediction accuracy. By learning from these examples, the multi-shot model can fine-tune its predictions based on the nuances present in the example data, leading to even more accurate results.

*3.0.3 Multi-Shot Prompting.* Multi-shot prompting using Gemini generally performed the best out of all three evaluation methods. For Participant 1, the multi-shot approach performed substantially better than the zero-shot approach for "Inspired" (0.58 ± 0.22) and "Alert" (0.38 ± 0.18), while achieving similar results for the other affects, as displayed in Table 1. For Participant 2, multi-shot predictions produced the highest accuracy for all affects except "Alert" (1.40 ± 0.34), as shown in Table 2. Notably, multi-shot prompting generally performed better than zero-shot prompting for the positive affects.

The enhanced performance of the multi-shot approach is influenced by its ability to learn from specific examples provided from the training dataset. By using these examples, the LLM gains a deeper understanding of the contextual nuances present in the data. This learning process allows the model to make more precise predictions, as a general model may not necessarily contain the knowledge to identify unique data relationships.

Additionally, providing the LLM with examples from the data creates a layer of personalisation that aims to enhance the understanding the intricacies of individual affective responses. Because human behaviour and affect varies between each individual, it can be challenging to predict and classify these correlations using a model that has been trained on a general corpus. Personalisation allows the model to adapt to the unique patterns and nuances of each individual's data. This approach recognises that affective responses are highly contextual and individualised [7], and by tailoring the model to account for these personal variations, we can achieve a deeper and more precise understanding of how screen text influences affective states. Therefore, we believe that incorporating personalised user data into predictive models is essential for tasks that involve complex and variable human affects, providing insights that are both more relevant and actionable.

## 4 FUTURE WORK AND LIMITATIONS

One limitation of our approach is the use of sentiment aggregates rather than analysing the full text content due to resource constraints. Future work should aim to incorporate the raw text content to extract their semantic meaning, which could provide richer and more detailed insights into how specific words, phrases, and contexts influence affective states. This approach could potentially reveal more nuanced relationships between screen text and affects, leading to improved prediction accuracy.

Another limitation of our study is that we initially employed a linear regression model to test the relationships between screen text sentiment and affect scores. While linear regression is a widely-used approach, it may have been more suitable to treat these scores as categorical data and use a specialised model, such as ordinal regression. While we do not expect significant changes to our overall results, these models may have potentially provided more reflective predictions of affective states as a baseline.

We also recommend exploring different training and evaluation windows to determine the optimal configuration for affect prediction. Our study used fixed durations, but it is likely that the ideal window size varies between participants due to differences in their behaviours and affective responses. By experimenting with various training and evaluation data sizes, future research can identify personalised optimal windows for each participant, enhancing the model's ability to make accurate predictions for each individual.

Additionally, there are numerous other ways to analyse screen text data that were not explored in this study. For instance, incorporating advanced natural language processing techniques such as topic modelling and named entity recognition could provide deeper insights into the content and its impact on affective states. Sentiment analysis could be combined with other linguistic features such as syntax, semantics, and pragmatics to create a more holistic understanding of the text's emotional influence. Furthermore, fine-tuning pre-trained LLMs on screen text data could further enhance their performance in predicting affects from screen text by allowing them to learn about the domain.

Table 1: Mean and standard deviation of MAE values for predicting each affect for Participant 1 across all three evaluation methods. Bold values represent the best-performing method for each affect.

| Affect | Linear Regression (Mean ± SD) | Gemini - Zero-Shot (Mean ± SD) | Gemini - Multi-Shot (Mean ± SD) |
| --- | --- | --- | --- |
| Active | 3.31 ± 1.12 | **0.80 ± 0.06** | 0.83 ± 0.10 |
| Determined | 1.65 ± 1.14 | 0.60 ± 0.09 | **0.58 ± 0.10** |
| Attentive | 1.92 ± 1.52 | 0.65 ± 0.24 | **0.63 ± 0.25** |
| Inspired | 4.78 ± 2.20 | 0.90 ± 0.15 | **0.58 ± 0.22** |
| Alert | 3.53 ± 2.41 | 1.10 ± 0.05 | **0.38 ± 0.18** |
| Upset | 3.54 ± 2.47 | **0.45 ± 0.10** | 0.55 ± 0.10 |
| Hostile | 3.17 ± 2.00 | **0.38 ± 0.08** | 0.45 ± 0.10 |
| Ashamed | 1.89 ± 1.14 | **0.13 ± 0.11** | 0.15 ± 0.12 |
| Nervous | 3.43 ± 3.30 | 0.95 ± 0.19 | **0.75 ± 0.25** |
| Afraid | 1.70 ± 1.14 | **0.08 ± 0.06** | **0.08 ± 0.06** |

Table 2: Mean and standard deviation of MAE values for predicting each affect for Participant 2 across all three evaluation methods. Bold values represent the best-performing method for each affect.

| Affect | Linear Regression (Mean ± SD) | Gemini - Zero-Shot (Mean ± SD) | Gemini - Multi-Shot (Mean ± SD) |
| --- | --- | --- | --- |
| Active | 2.47 ± 0.84 | 1.70 ± 0.10 | **0.33 ± 0.10** |
| Determined | 2.51 ± 2.04 | 1.73 ± 0.15 | **0.30 ± 0.10** |
| Attentive | 2.10 ± 0.94 | 1.53 ± 0.18 | **0.50 ± 0.16** |
| Inspired | 1.56 ± 0.77 | 2.40 ± 0.22 | **0.78 ± 0.22** |
| Alert | 2.62 ± 0.73 | **1.20 ± 0.13** | 1.40 ± 0.34 |
| Upset | 2.37 ± 0.63 | 1.98 ± 0.30 | **1.28 ± 0.20** |
| Hostile | 2.98 ± 1.51 | 1.28 ± 0.24 | **1.18 ± 0.30** |
| Ashamed | 2.68 ± 0.80 | 1.88 ± 0.24 | **1.28 ± 0.22** |
| Nervous | 1.98 ± 1.05 | 2.90 ± 0.22 | **0.65 ± 0.31** |
| Afraid | 3.44 ± 4.14 | 3.53 ± 0.28 | **0.68 ± 0.13** |

## 5 CONCLUSION

Our study demonstrates that it is feasible to predict affective states based on screen text sentiment. By leveraging advanced NLP techniques, particularly multi-shot prompting with LLMs, we achieved a significant improvement in prediction accuracy compared to traditional methods such as linear regression. The linear regression method exhibited the lowest accuracy in predicting affective states, primarily due to its simplistic assumption of a linear relationship between screen text sentiment and affects, which fails to capture the complex and nuanced interactions inherent in human emotional responses. In contrast, the zero-shot approach, which utilises the LLM's pre-trained knowledge, provided better predictions but was less accurate than the multi-shot method. Multi-shot prompting provided the most accurate predictions by using specific examples from the training data. This approach produced lower error and higher predictive accuracy, indicating the model could potentially understand the context and nuance of the data more effectively. By incorporating detailed examples, the multi-shot method could capture more of the patterns and relationships between screen text and affective states, which are often missed by less sophisticated models.

However, it is important to note that the accuracy of affect prediction varied across different individuals. This variability can be attributed to the unique nature of each person's affective responses and the contextual factors influencing them. Human affects are complex and influenced by a myriad of personal experiences and context outside of their digital life, making it challenging to develop a one-size-fits-all predictive model. Personalisation, as achieved through multi-shot prompting, helps to mitigate this issue by tailoring the model to the specific nuances and patterns in an individual's data, thereby enhancing prediction accuracy.

Overall, our findings show that the rich context that screen text data provides can be utilised alongside personalised models for affect prediction tasks. By leveraging the contextual understanding and adaptability of LLMs, particularly through multi-shot prompting, we can achieve more accurate and meaningful insights into how screen text influences affective states. This approach not only improves the reliability of predictions but also provides a deeper understanding of individual variations in emotional responses, making it a valuable tool for future research and practical applications in wellbeing and beyond.

## ACKNOWLEDGMENTS

This work is partially funded by NHMRC grants 1170937 and 2004316, AUSMURI grant 13203896, and CISCO grant 2021-2327463696.

## REFERENCES

[1] Sofian Berrouiguet, David Ramírez, María Luisa Barrigón, Pablo Moreno-Muñoz, Rodrigo Carmona Camacho, Enrique Baca-García, and Antonio Artés-Rodríguez. 2018. Combining Continuous Smartphone Native Sensors Data Capture and Unsupervised Data Mining Techniques for Behavioral Changes Detection: A Case Series of the Evidence-Based Behavior (eB2) Study. *JMIR mHealth and uHealth* 6, 12 (Dec. 2018), e197. https://doi.org/10.2196/mhealth.9472
[2] Simon D'Alfonso and Tianyi Zhang. 2024. StudentSense. (2024). https://doi.org/10.17605/OSF.IO/DJS2Y
[3] Munmun De Choudhury, Michael Gamon, Scott Counts, and Eric Horvitz. 2021. Predicting Depression via Social Media. *Proceedings of the International AAAI Conference on Web and Social Media* 7, 1 (Aug. 2021), 128–137. https://doi.org/10.1609/icwsm.v7i1.14432

[4] Denzil Ferreira, Vassilis Kostakos, and Anind K. Dey. 2015. AWARE: Mobile Context Instrumentation Framework. *Frontiers in ICT* 2 (April 2015). https://doi.org/10.3389/fict.2015.00006

[5] Mads Frost, Gabriela Marcu, Rene Hansen, Karoly Szaántó, and Jakob Bardram. 2011. The MONARCA Self-assessment System: Persuasive Personal Monitoring for Bipolar Patients. In *Proceedings of the 5th International ICST Conference on Pervasive Computing Technologies for Healthcare (PERVASIVEHEALTH)*. IEEE. https://doi.org/10.4108/icst.pervasivehealth.2011.246050

[6] Gemini Team. 2024. Gemini 1.5: Unlocking multimodal understanding across millions of tokens of context. https://doi.org/10.48550/ARXIV.2403.05530

[7] Katharine H. Greenaway, Elise K. Kalokerinos, and Lisa A. Williams. 2018. Context is Everything (in Emotion Research). *Social and Personality Psychology Compass* 12, 6 (May 2018). https://doi.org/10.1111/spc3.12393

[8] Akshay Kulkarni and Adarsha Shivananda. 2021. *Natural Language Processing Recipes: Unlocking Text Data with Machine Learning and Deep Learning Using Python*. Apress. https://doi.org/10.1007/978-1-4842-7351-7

[9] Pranav Kulkarni, Reuben Kirkham, and Roisin McNaney. 2022. Opportunities for Smartphone Sensing in E-Health Research: A Narrative Review. *Sensors* 22, 10 (May 2022), 3893. https://doi.org/10.3390/s22103893

[10] Laura Montanini, Nicola Sabino, Susanna Spinsante, and Ennio Gambi. 2018. Smartphone as unobtrusive sensor for real-time sleep recognition. In *2018 IEEE International Conference on Consumer Electronics (ICCE)*. IEEE. https://doi.org/10.1109/icce.2018.8326220

[11] Byron Reeves, Thomas Robinson, and Nilam Ram. 2020. Time for the Human Screenome Project. *Nature* 577, 7790 (Jan. 2020), 314–317. https://doi.org/10.1038/d41586-020-00032-5

[12] Ensar Arif Sağbaş, Serdar Korukoglu, and Serkan Balli. 2020. Stress Detection via Keyboard Typing Behaviors by Using Smartphone Sensors and Machine Learning Techniques. *Journal of Medical Systems* 44, 4 (Feb. 2020). https://doi.org/10.1007/s10916-020-1530-z

[13] Victor Sanh, Lysandre Debut, Julien Chaumond, and Thomas Wolf. 2019. DistilBERT, a distilled version of BERT: smaller, faster, cheaper and lighter. https://doi.org/10.48550/ARXIV.1910.01108

[14] Shams Shapsough, Ahmed Hesham, Youssef Elkhorazaty, Imran A. Zualkernan, and Fadi Aloul. 2016. Emotion recognition using mobile phones. In *2016 IEEE 18th International Conference on e-Health Networking, Applications and Services (Healthcom)*. IEEE. https://doi.org/10.1109/healthcom.2016.7749470

[15] Sonish Sivarajkumar, Mark Kelley, Alyssa Samolyk-Mazzanti, Shyam Visweswaran, and Yanshan Wang. 2024. An Empirical Evaluation of Prompting Strategies for Large Language Models in Zero-Shot Clinical Natural Language Processing: Algorithm Development and Validation Study. *JMIR Medical Informatics* 12 (April 2024), e55318. https://doi.org/10.2196/55318

[16] Caitlin A. Stamatis, Jonah Meyerhoff, Tingting Liu, Garrick Sherman, Harry Wang, Tony Liu, Brenda Curtis, Lyle H. Ungar, and David C. Mohr. 2022. Prospective associations of text-message-based sentiment with symptoms of depression, generalized anxiety, and social anxiety. *Depression and Anxiety* 39, 12 (Oct. 2022), 794–804. https://doi.org/10.1002/da.23286

[17] Michael F. Steger and Todd B. Kashdan. 2009. Depression and everyday social activity, belonging, and well-being. *Journal of Counseling Psychology* 56, 2 (2009), 289–300. https://doi.org/10.1037/a0015416

[18] Songyan Teng, Simon D'Alfonso, and Vassilis Kostakos. 2024. A Tool for Capturing Smartphone Screen Text. In *Proceedings of the CHI Conference on Human Factors in Computing Systems (CHI '24)*. ACM. https://doi.org/10.1145/3613904.3642347

[19] Edmund R. Thompson. 2007. Development and Validation of an Internationally Reliable Short-Form of the Positive and Negative Affect Schedule (PANAS). *Journal of Cross-Cultural Psychology* 38, 2 (March 2007), 227–242. https://doi.org/10.1177/0022022106297301

[20] Ana-Sabina Uban, Berta Chulvi, and Paolo Rosso. 2021. An emotion and cognitive based analysis of mental health disorders from social media data. *Future Generation Computer Systems* 124 (Nov. 2021), 480–494. https://doi.org/10.1016/j.future.2021.05.032

[21] Niels van Berkel, Simon D'Alfonso, Rio Kurnia Susanto, Denzil Ferreira, and Vassilis Kostakos. 2022. AWARE-Light: a smartphone tool for experience sampling and digital phenotyping. *Personal and Ubiquitous Computing* 27, 2 (Nov. 2022), 435–445. https://doi.org/10.1007/s00779-022-01697-7

[22] Rui Wang, Fanglin Chen, Zhenyu Chen, Tianxing Li, Gabriella Harari, Stefanie Tignor, Xia Zhou, Dror Ben-Zeev, and Andrew T. Campbell. 2014. StudentLife: Assessing Mental Health, Academic Performance and Behavioral Trends of College Students using Smartphones. In *Proceedings of the 2014 ACM International Joint Conference on Pervasive and Ubiquitous Computing*. ACM. https://doi.org/10.1145/2632048.2632054

[23] Hua-Cong Yang, Yi-Chao Li, Zhi-Yu Liu, and Jie Qiu. 2014. HARLib: A human activity recognition library on Android. In *2014 11th International Computer Conference on Wavelet Actiev Media Technology and Information Processing(ICCWAMTIP)*. IEEE. https://doi.org/10.1109/iccwamtip.2014.7073416